\documentclass[aps,prd,preprintnumbers,showpacs,showkeys,nofootinbib,
superscriptaddress,fleqn,floatfix,tightenlines,10pt,
byrevtex]{revtex4-2}
\usepackage{amsmath,amsfonts,amssymb,amscd,amsxtra,amsthm}
\usepackage{appendix}
\usepackage[unicode]{hyperref}
\usepackage{graphicx,subfigure}
\graphicspath{{./Figures/}}
\usepackage{indentfirst}
\usepackage{float}
\usepackage{bm}
\usepackage{physics}
\usepackage{multirow}
\usepackage{makecell}
\usepackage{ltablex}
\usepackage{siunitx}

\usepackage[utf8]{inputenc}
\usepackage[dvipsnames]{xcolor} 
\graphicspath{{./Figures/}}
\usepackage{indentfirst}
\usepackage{float}

\usepackage[dvipsnames]{xcolor}
\usepackage{umoline}

\newcommand{\Slash}[1]{\ooalign{\hfil/\hfil\crcr$#1$}}
\makeatletter

\begin{document}
\preprint{INHA-NTG-06/2026}
\title{Heavy mesons from the QCD instanton vacuum beyond the
static limit}

\author{Ki-Hoon Hong}
\email{kihoon.hong@inha.ac.kr}
\affiliation{Institute of Quantum Science, Inha University, Incheon 22212,
Republic of Korea}

\author{Yongwoo Choi}
\email{ywchoi@hoseo.edu}
\affiliation{School of AIT, Hoseo University, Asan 31499, Republic of Korea}
\affiliation{Quantum AI Innovative Technology Institute (Q-AIT), Hoseo University, Asan 31499, Republic of Korea}

\author{Nurmukhammad Rakhimov}
\email{n.r.rakhimov@gmail.com}
\affiliation{Institute of Quantum Science, Inha University, Incheon 22212,
Republic of Korea}

\author{Hyun-Chul Kim}
\email{hchkim@inha.ac.kr}
\affiliation{Department of Physics, Inha University, Incheon 22212,
Republic of Korea}
\affiliation{Institute of Quantum Science, Inha University, Incheon 22212,
Republic of Korea}
\affiliation{School of Physics, Korea Institute for Advanced Study
(KIAS), Seoul 02455, Republic of Korea}

\begin{abstract}
We study pseudoscalar heavy mesons in the QCD instanton vacuum beyond
the static limit. Finite-mass effects in the heavy-light loop are
encoded in a separable effective vertex built from a profile function
$\phi(\vec{p})$, kept distinct from the static Wilson-line form factor
$F_Q^{(\infty)}(\vec{q})$ of the $m_Q\to\infty$ limit. The
pseudoscalar two-point function fixes the residual mass $\Lambda$ and
the residue-normalized meson-quark coupling, from which we
{evaluate the decay constant and the spin-independent
  kinetic matrix element including the retained $1/m_Q$ derivative and
  kinetic contributions. The Isgur-Wise function is evaluated
  separately in the static heavy-quark limit, and its zero-recoil
  slope is extracted by using the same residue-normalized heavy-light
  vertex}. The subleading calculation is restricted to the kinetic
(derivative) part of the HQET operators. For a representative vertex,
{calibrated at fixed $m_b^{\mathrm{eff}}=5.04\,\mathrm{GeV}$
  to the leading-order decay constant
  $f_B^{\mathrm{LO}}=180.0\,\mathrm{MeV}$ and the spin-averaged mass
  $\bar m_B=5.313\,\mathrm{GeV}$}, we 
obtain { $\delta f_B^{(1/m_Q,\mathrm{partial})}=+121.9$~MeV and
  $f_B^{\mathrm{partial}}=301.9$~MeV},
$\Lambda=\,{190.7}$~MeV,
$\lambda_1^{(\partial)}=-\,{0.773}\,\mathrm{GeV}^2$, and
$\rho_{\mathrm{IW}}^2=\,{1.153}$. The kinetic contribution
yields a mass shift of order $\Lambda/2$ and a  {sizable
  partial $1/m_Q$ correction}, indicating that the spin-independent
nonperturbative $1/m_Q$ sector is a sensitive probe of the finite-mass
heavy-light vertex. 
\end{abstract}

\date{\today}
\maketitle

\section{Introduction}\label{sec:intro}
Heavy mesons containing a single heavy quark provide a standard system
for studying how quantum chromodynamics (QCD) separates the
short-distance physics at the scale $m_Q$ from the nonperturbative
dynamics at hadronic scales. In the heavy-quark mass limit
$m_Q\to\infty$, the spin of the heavy quark decouples from the gluon
field, so that the heavy-quark spin and the total angular momentum of
the light degrees of freedom are separately conserved. The dynamics also
become independent of the heavy-quark flavor, since the heavy quark
enters only as a static color source. These properties constitute the
heavy-quark spin-flavor symmetry~\cite{Isgur:1989vq, Isgur:1990yhj,
Georgi:1990um, Georgi:1991mr, Wise:1993wa}. In this limit, the internal
dynamics of a heavy hadron is governed entirely by the light degrees of
freedom. Because $m_Q$ is much larger than the light-quark masses and the
typical hadronic scale, $1/m_Q$ serves as a small expansion parameter. A
systematic expansion of the heavy-quark QCD Lagrangian in $1/m_Q$ yields
heavy-quark effective theory (HQET)~\cite{Neubert:1993mb,Shifman:1995dn,
Neubert:1996wg, Casalbuoni:1996pg}.

In HQET, the perturbative contributions to matrix elements of heavy
hadrons are contained in the Wilson coefficients, whereas the
long-distance contributions must be supplied by nonperturbative
methods~\cite{Neubert:1993mb, Shifman:1995dn, Neubert:1996wg,
Casalbuoni:1996pg}. The QCD instanton vacuum provides one such
framework for computing these nonperturbative contributions.
It realizes the spontaneous breakdown of chiral symmetry (SB$\chi$S) in
QCD~\cite{Shuryak:1981ff, Diakonov:1983hh, Diakonov:1985eg} (see also
the reviews~\cite{Schafer:1996wv, Diakonov:2002fq}).
The mechanism of the SB$\chi$S can be explained as follows: a quark 
propagates through the instanton medium, which consists of
instantons ($I$) and antiinstantons ($\bar{I}$). As the quark moves 
between random $I$'s and $\bar{I}$'s, its helicity undergoes a
change due to the different handedness of the $I$ and $\bar{I}$ zero
modes. This process yields the SB$\chi$S, and the 
quark acquires a dynamical quark mass, $M(k)$, which is
momentum-dependent. $M(k)$ comes from the 
Fourier transform of the fermionic zero mode~\cite{Diakonov:1985eg,
  Diakonov:2002fq}. The value at zero virtuality, $M(0)\sim
350\,\mathrm{MeV}$, is determined by the saddle-point equation.
The effective low-energy QCD partition function derived from the
instanton vacuum has been successfully employed to describe low-lying
light hadrons~\cite{Diakonov:1987ty, Musakhanov:2002xa, Nam:2007gf,
  Nam:2007fx, Nam:2011yw, Son:2015bwa, Shim:2017wcq, Shim:2018rwv, 
  Christov:1995vm, Diakonov:1997sj}.

This theoretical framework was subsequently extended
to systems that include heavy quarks~\cite{Diakonov:1989un,
  Chernyshev:1995gj, Yakhshiev:2016keg, Yakhshiev:2018juj,
  Musakhanov:2020hvk, Hong:2022sht, Hong:2024ptu}.
A heavy quark interacts with the background gauge field through a Wilson
line, and its coupling to the instanton-antiinstanton ensemble fixes the
instanton contributions to the heavy-quark mass and the heavy-quark
potential. Combining the heavy quark with the light quarks yields an
effective heavy-light quark interaction. This interaction is nonlocal, so
that the light and heavy quarks each carry a form factor: the light-quark
form factor $F(p)$ generates the momentum-dependent dynamical
mass~\cite{Diakonov:1985eg}, while the Wilson line gives the static
heavy-line form factor $F_Q^{(\infty)}(\vec{q})$ in the $m_Q\to\infty$
limit~\cite{Hong:2024ptu}.

In our previous work~\cite{Hong:2024ptu}, the momentum dependence of the
heavy-line form factor was neglected by setting
$F_Q^{(\infty)}(\vec q)\to F_Q^{(\infty)}(\vec 0)=1$. The normalized
static Wilson-line form factor $F_Q^{(\infty)}(\vec q)$ describes the
momentum transfer along the heavy line in the $m_Q\to\infty$ limit. The
present bound-state loop calculation, however, requires the nonlocal
heavy-light vertex as a function of the relative three-momenta at the two
heavy-light vertices. We therefore represent the finite-mass heavy-light
kernel by the separable effective form
\begin{align}
\mathcal{F}(\vec p,\vec l;m_Q)\simeq
\frac{\phi(\vec p)\phi(\vec l)}{N_\phi^2},
\end{align}
{where $\phi(\vec p)$ denotes the finite-mass heavy-light vertex
  profile and $N_\phi$ stands for a momentum-independent normalization
  factor. The two heavy-line functions play different roles:
  $F_Q^{(\infty)}(\vec q)$ depends on the momentum transfer along
  the static heavy line, whereas $\phi(\vec p)$ depends on the
  relative momentum at a heavy-light vertex. The pole condition and
  its residue fix the interaction strength used below.
We then calculate the residual mass $\Lambda$, the pseudoscalar decay
  constant, the spin-independent kinetic matrix element, and the
  Isgur-Wise slope for a single light flavor. The explicit $1/m_Q$
  calculation is limited to the ordinary-derivative part of the
  subleading current and kinetic insertion. Gauge-field terms in the
  covariant derivatives and the chromomagnetic operator require the
  corresponding effective gluonic operators and are not included here.}

{We sketch the structure of the present work. Section~\ref{sec:1}
  provides the nonlocal heavy-light interaction, 
  its separable form, and the pole-residue
  normalization. Sections~\ref{sec:decay} and 
\ref{sec:kinetic} contain the decay-constant and kinetic calculations,
respectively. The Isgur-Wise function is discussed in Section~\ref{sec:iw},
and the numerical analysis is given in Section~\ref{sec:numerics}.}

\section{Heavy-light interaction from the instanton vacuum}
\label{sec:1}
We start from the effective heavy-light quark action for the light
flavor number $N_f=1$ derived from the
instanton vacuum~\cite{Hong:2024ptu}:
\begin{align}
    S_{\rm eff} &= S_0 + S_{\text{int}},\label{eq:S_eff}\\
    S_0 ~ &= {-}\int d^4x~\left[
          q^\dagger(x)\left(i\Slash{\partial}+iM(i\partial)\right)q(x)
          + 
          h^\dagger(x)(iv\cdot\partial)h(x) \right],\label{eq:S_0}\\ 
    S_{\text{int}} &= {-}M_q\int
    d^4z\int\frac{d^4p_q}{(2\pi)^4}
    \frac{d^4l_q}{(2\pi)^4}~ e^{-i(p_q-l_q)\cdot z}F(p_q)F(l_q) 
    \int d^4x \,d^4y~ \frac{1}{N_c}\tr_c \langle
                     x|\theta^{-1}(w-\theta)\theta^{-1}|y\rangle\cr 
    &\quad\times \left[-\left(q^\dagger(p_q)q(l_q)\right)
      \left(h^\dagger(x) h(y)\right)  
    +\frac{1}{8}\left(h^\dagger(x)\Gamma_i
      q(l_q)\right)\,\left(q^\dagger(p_q)\Gamma_i h(y)\right)\right], \label{eq:S_int}
\end{align}
where
$\Gamma_i=\{\bm{1},\,\gamma_5,\,\gamma_\mu,\,i\gamma_\mu\gamma_5,\,
\sigma_{\mu\nu}/\sqrt{2}\}$, $S_0$ and $S_{\mathrm{int}}$ represent
the free and interaction actions, respectively.
The fields $q$ and $h$ denote the corresponding light and heavy quark
fields; a heavy-flavor label can be attached to $h$ when
needed. $M(i\partial)$ stands for the dynamical quark mass for
the light quark, $M(k)=M_q F^2(k)$ with $M_q =
345\,\mathrm{MeV}$~\cite{Diakonov:1985eg,Diakonov:2002fq}, where
$F(k)$ is given by
\begin{align}
F(k) = -z \frac{d}{dz} \left[I_0(z) K_0(z) - I_1(z)
  K_1(z)\right]  \Big|_{z=\bar{\rho}{k}/2},
\end{align}
where $I_{n}(z)$ and $K_n(z)$ are
the modified Bessel functions of the first and second kinds of order
$n$, respectively, and $\bar{\rho}$ is the average instanton size.
In the heavy-quark mass limit $m_Q\to \infty$, the action exhibits
heavy-quark spin-flavor
symmetry~\cite{Georgi:1991mr,Wise:1993wa,Neubert:1993mb}.
The interaction term in Eq.~\eqref{eq:S_int} is given as a nonlocal
four-point interaction between heavy-light quarks, where the quark
form factor $F(k)$ is attached to each light-quark, whereas heavy
quarks are dressed by a Wilson-line form factor~\cite{Diakonov:1989un,
Hong:2024ptu}. Here $\theta^{-1}\,(=d/dt)$ is the free static-heavy-quark propagator,
and $w=(\theta^{-1}-iA_{I\mu}\dot{x}_\mu(t))^{-1}$ is the corresponding
propagator in the field of a single (anti-)instanton~\cite{Diakonov:1989un}.
The notation $(q^\dagger q) = q_{ai}^\dagger q^{ai}$ indicates
contraction over spin and color indices, and the four-fermion terms
arise from Fierz reordering in color and spin spaces.
{The overall signs in Eqs.~\eqref{eq:S_0} and \eqref{eq:S_int}
follow from the Wick rotation of the Minkowski action to the Euclidean
one used throughout this work. The propagators and vertices below
are defined consistently with this convention.}
Details of the static Wilson-line kernel entering the heavy-light
interaction are provided in Appendix~\ref{app:A}. In the HQET limit,
retaining only the leading contribution in the residual energies
$\omega_{1,2}\sim\Lambda_{\rm QCD}$ yields a normalized momentum-space
form factor $F_Q^{(\infty)}(\vec q)$, which characterizes the momentum
dependence of the static reference kernel. Using this result, the
heavy-light interaction can be written as
\begin{align}
    S_{qQ} &= {+}g^2\int
             \frac{d^4p_q}{(2\pi)^{4}}\frac{d^4l_q}{(2\pi)^{4}}
             \frac{d^4k_1}{(2\pi)^{4}}\frac{d^4k_2}{(2\pi)^{4}}~
             (2\pi)^{4}\delta^{(4)}(p_q-l_q+k_1-k_2)
             F(p_q)F(l_q)F_Q^{(\infty)}(\vec{k}_1-\vec{k}_2)\cr 
    &\quad\times \sum_i \left[ q^\dagger(-p_q)\Gamma_ih(k_1) \right]
    \left[ h^\dagger(k_2)\Gamma_i q(-l_q) \right],\label{eq:heavy-light_int} 
\end{align}
where the coupling constant is defined as $g^2 \equiv M_q\Delta M_Q / (16n)$, and the momentum-dependent dynamical quark mass is given by $M(p_q) = \frac{(2\pi\bar{\rho})^2\lambda}{N_c}F^2(p_q) \equiv M_q F^2(p_q)$. Here, we redefine $T(\vec{q})\equiv -\frac{\Delta
M_Q}{2n}F_Q^{(\infty)}(\vec q=\vec k_1-\vec k_2)$, using the
instanton contribution $\Delta M_Q=\frac{16\pi n\bar{\rho}^3}{N_c}\int_0^\infty
d\tilde r ~\tilde r^2 F_Q^{(\infty)}(\tilde r) = 68\,$MeV to the heavy quark
mass and the instanton density $n=N/V=(200\,\rm MeV)^4$:
\begin{align}
F_Q^{(\infty)}(\vec{q}) &= \frac{\int_0^\infty d\tilde r~ \tilde r^2
    j_0(\bar{\rho}|\vec q|\tilde{r})
    F_Q^{(\infty)}(\tilde r)}{\int_0^\infty d\tilde r~
    \tilde r^2F_Q^{(\infty)}(\tilde r)} \label{eq:FQ_inf_mom}\\
F_Q^{(\infty)}(\tilde r) &= \cos^2
    \left(\frac{\pi \tilde r}{2\sqrt{\tilde r^2+1}}\right).\label{eq:FQ_inf_pos}
\end{align}
In Ref.~\cite{Hong:2024ptu}, this momentum dependence was replaced by
the static momentum approximation, $F_Q^{(\infty)}(|\vec q|\to0)=1$. The
normalization in Eq.~\eqref{eq:FQ_inf_mom} is chosen such that
$F_Q^{(\infty)}(\vec q=0)=1$, so the Wilson-line form factor reduces to
the static result in the forward limit. The position-space function
$F_Q^{(\infty)}(\tilde r)$ is fixed by the Wilson-line structure in the
instanton background and depends only on $\tilde r=r/\bar\rho$.

The overall strength $\Delta M_Q$ is factored out, so
$F_Q^{(\infty)}(\vec q)$ describes only the momentum dependence generated
by the Wilson line in the large-$m_Q$ limit. Since this form factor is
a function of the momentum transfer $\vec q$ along the heavy line, we do
not identify it directly with the heavy-light vertex needed in the
bound-state loop calculation. Instead, we replace the direct use of
$F_Q^{(\infty)}(\vec q)$ by the phenomenological separable ansatz
\begin{align}
\mathcal{F}(\vec p,\vec l;m_Q)\simeq
\frac{\phi(\vec p)\phi(\vec l)}{N_\phi^2},
\label{eq:FQ_separable}
\end{align}
where $\vec p$ and $\vec l$ are the relative three-momenta at the two
heavy-light vertices. Here $\phi(\vec p)$ is the heavy-light vertex
profile and $N_\phi$ is its dimensionless overall normalization factor.
This ansatz supplies the vertex momentum dependence used in the
calculation of the residual mass, decay constant, kinetic correction,
and Isgur-Wise function.

In the convention used here, the normalized $F_Q^{(\infty)}(\vec q)$,
the effective finite-mass kernel $\mathcal{F}(\vec p,\vec l;m_Q)$, and
$\phi(\vec p)$ are dimensionless. Imposing the ansatz in
Eq.~\eqref{eq:FQ_separable}, one can rewrite the interaction term in
Eq.~\eqref{eq:heavy-light_int} in terms of heavy--light currents as follows:
\begin{align}
S_{qQ}^{\rm sep}
    &= {+}G_0^2\sum_i\int\frac{d^4 p_H}{(2\pi)^4}~
    J_i(p_H)J_i^\dagger(p_H),
    \label{eq:SqQ_sep}\\
J_i(p_H)
    &\equiv \int\frac{d^4p}{(2\pi)^4}~
    q^\dagger(-p)\Gamma_i F(p)\phi(\vec p)h(p_H-p),
    \label{eq:J_sep}\\
J_i^\dagger(p_H)
    &\equiv \int\frac{d^4l}{(2\pi)^4}~
    h^\dagger(p_H-l)\Gamma_i F(l)\phi(\vec l)q(-l),
    \label{eq:Jdag_sep}
\end{align}
where $S_{qQ}^{\rm sep}$ follows from Eq.~\eqref{eq:heavy-light_int} after applying
the separable ansatz in Eq.~\eqref{eq:FQ_separable}, integrating over
$k_2$, and redefining $p_H=p_q+k_1$, $p_q=p$, and $l_q=l$.
The heavy-light current entering the bosonized theory is therefore
defined by a single four-momentum integral over the relative momentum.
The finite-mass heavy-line effect is encoded in the rest-frame vertex
profile $\phi(\vec p)$. Before meson-field residue normalization, the
separable interaction has the coefficient $G_0^2=g^2/N_\phi^2$,
($[G_0]=[g]$).

The separable interaction can be bosonized as:
\begin{align}
S_{qQ\Phi_v}^{\rm sep}
    =\sum_i\int\frac{d^4 p_H}{(2\pi)^4}~
    \left[\Phi_{iv}^\dagger(p_H)\Phi_{iv}(p_H)
    -G_0\left\{\Phi_{iv}^\dagger(p_H)J_i(p_H)
    {-}J_i^\dagger(p_H)\Phi_{iv}(p_H)\right\}\right].
\end{align}
Here, the index $i$ specifies the meson channel and the subscript $v$
indicates the HQET velocity sector. The relative-momentum dependence
resides in the current $J_i(p_H)$ through the factors $F(p)$ and
$\phi(\vec p)$, so the quadratic Hubbard-Stratonovich term is diagonal
in the residual momentum of the heavy meson $p_H$. The dimensional
rescaling associated with the physical heavy-meson residue is
performed only after the pole residue is extracted below.

Using the light- and heavy-quark Lagrangian in Eq.~\eqref{eq:S_0} and
integrating out the quark fields, we obtain the quadratic effective
action for the physical HQET meson field in terms of the heavy-meson
residual momentum $p_H$,
\begin{align}
S_{\rm eff}^{(2)}
    =\sum_i\int\frac{d^4 p_H}{(2\pi)^4}~
    \Phi_{iv}^\dagger(p_H)
    \left[1-G_0^2\Sigma_i(p_H)\right]
    \Phi_{iv}(p_H),
\end{align}
where the heavy-light loop function $\Sigma_i$ is given by
\begin{align}
\Sigma_i(p_H)
    =\int\frac{d^4p}{(2\pi)^4}~|\phi(\vec p)|^2F^2(p)
    \tr_D\left[S_q(-p)\Gamma_iS_h(p_H-p)\Gamma_i\right].
\end{align}
Here, $p=(\vec p,\,p_4)$ represents the internal light-quark loop
momentum. In our convention, the light- and heavy-quark propagators
are given by
\begin{align}
    S_q(p) = \frac{\Slash{p}+iM(p)}{{p}^2+M^2(p)},\quad S_h(k) = \frac{1+\Slash{v}}{2v\cdot k}.
    \label{eq:propagators}
\end{align}
Due to the structure of the HQET propagator, the dependence of the loop
function on the residual momentum reduces to the scalar form $v \cdot p_H$,
which directly determines the inverse meson propagator,
\begin{align}
    S_{H,i}^{-1}(v\cdot p_H)=1-G_0^2 \, \Sigma_i(v\cdot p_H).
\end{align}
A physical heavy meson is obtained when this inverse propagator
vanishes.
In Euclidean space, we set $p_H=i \Lambda v$, so that the pole
position defines the residual-mass parameter $\Lambda$ in the present
effective theory.

For the pseudoscalar channel, $\Gamma_P=\gamma_5$, the pole condition
is given by $G_0^2\,\Sigma_P(i\Lambda)=1$.
By exploiting the reflection symmetry of the $p_4$ integration domain,
the odd terms in $p_4$ vanish, and the remaining even part of
the integrand can be written as
\begin{align}
\mathcal{I}_P(p;\Lambda)
    =\frac{F^2(p)}{\Lambda^2+p_4^2}
    \frac{2M(p)\Lambda-2p_4^2}
    {|\vec p|^2+p_4^2+M^2(p)},
\end{align}
where $F(p)$ and $M(p)$ denote
$F(\sqrt{|\vec p|^2+p_4^2})$ and $M(\sqrt{|\vec p|^2+p_4^2})$,
respectively. We then define
\begin{align}
    \mu_P(i\Lambda,\, p_4)
    \equiv \int\frac{d^3p}{(2\pi)^3}~|\phi(\vec p)|^2\,
    \mathcal{I}_P(p;\, \Lambda),
\end{align}
so that
\begin{align}
    \Sigma_P(i\Lambda)=\int_0^\infty\frac{dp_4}{\pi}~
    \mu_P(i\Lambda,\, p_4).
\end{align}
The pole residue fixes the dimensional rescaling from the bosonized
field $\Phi_{Pv}$ to the canonically normalized HQET meson field. Near
$v\cdot p_H=i\Lambda$, the inverse propagator behaves as
\begin{align}
S_H^{-1}
    =1-G_0^2 \,\Sigma_P(v\cdot p_H)
    \simeq -G_0^2 \,\Sigma'_P(i\Lambda)(v\cdot p_H-i\Lambda),
    \label{eq:inv_meson_prop}
\end{align}
where the prime denotes the derivative with respect to $v\cdot p_H$ at
the pole. Since this derivative is taken with respect to a dimensionful
variable, the combination $iG_0^2 \,\Sigma'_P(i\Lambda)$ has mass dimension
$M^{-1}$. We define the pole-residue factor
\begin{align}
    \mathcal{R}_H\equiv iG_0^2 \,\Sigma'_P(i\Lambda),
    \quad [\mathcal{R}_H]=M^{-1}.
\end{align}

{The Hubbard--Stratonovich kernel is dimensionless. A physical
  HQET inverse propagator, by contrast, is linear in the residual
  energy, so the pole residue $\mathcal{R}_H$ supplies the dimensional
  rescaling as follow:} 
\begin{align}
    H_v(p_H)=\sqrt{\frac{\mathcal{R}_H}{2}}\,\Phi_{Pv}(p_H),
    \quad\Phi_{Pv}(p_H)=\sqrt{\frac{2}{\mathcal{R}_H}}\,H_v(p_H).
\end{align}
The residue-normalized heavy-light coupling is then defined as
\begin{align}
    G\equiv \sqrt{\frac{2}{\mathcal{R}_H}}\,G_0.
    \label{eq:G_rescaled}
\end{align}

{Only the combination $G_0=g/N_\phi$ survives in the separable
  interaction; $N_\phi$ itself need not be chosen. Once the profile
  and pole position are specified, the pole condition fixes $G_0$. For
  the field normalization we set
  $\mathcal{R}_H=2\,\mathrm{GeV}^{-1}$. This convention does not add
  another interaction strength or observable. It also gives
  $[G]=M^{-1/2}$ from $[G_0]=[\mathcal{R}_H]=M^{-1}$.}
With this definition, one has
\begin{align}
    iG^2\, \Sigma'_P(i\Lambda)=2.
    \label{eq:compositeness_G}
\end{align}
Explicitly,
\begin{align}
\Sigma'_P(i\Lambda)
    = -i\int_0^\infty\frac{dp_4}{\pi}~\frac{\partial\mu_P(i\Lambda,p_4)}{\partial\Lambda}
    = -i\int_0^\infty\frac{dp_4}{\pi}
    \int\frac{d^3p}{(2\pi)^3}~
    \frac{|\phi(\vec p)|^2F^2(p)}{(\Lambda^2+p_4^2)^2}
    \frac{4p_4^2\Lambda+2M(p)(p_4^2-\Lambda^2)}
    {|\vec p|^2+p_4^2+M^2(p)}.
\end{align}

{Using the pole condition, we give the pre-rescaling quantities}
\begin{align}
G_0^2=\frac{1}{\Sigma_P(i\Lambda)},
    \quad\mathcal{R}_H=i\frac{\Sigma'_P(i\Lambda)}{\Sigma_P(i\Lambda)},
\end{align}
whereas the residue-normalized coupling entering external meson matrix
elements is 
\begin{align}
    G^2=\frac{2}{i\Sigma'_P(i\Lambda)}.
\end{align}
{The heavy-light vertex for the matrix elements is given by}
\begin{align}
    \mathcal{V}_{H,i}(p)=\frac{G}{\sqrt{N_c}}F(p)\phi(\vec
  p)\Gamma_i,\
  {\mathcal{V}_{H^\dagger,i}(p)=-\frac{G}{\sqrt{N_c}}F(p)\phi(\vec
  p)\Gamma_i}. 
    \label{eq:heavy-light_vertex}
\end{align}
Each internal quark line carries the four-momentum integration with
momentum conservation $(2\pi)^4\delta^{(4)}(p_H-p_q-k)$ at the
vertex. The pole condition is written in terms of $G_0$, while decay
constants, kinetic matrix elements, and heavy-to-heavy form factors are
written in terms of the residue-normalized coupling $G$.

\section{Heavy-light current and weak decay constants}
\label{sec:decay}

{For a fixed profile, the pseudoscalar two-point function determines
$\Lambda$ and the residue-normalized heavy-light vertex. These quantities
are then used in the matrix element of the axial-vector current that
defines the pseudoscalar decay constant. We restrict the explicit
calculation to the pseudoscalar channel. Note that vector states and
hyperfine observables require the chromomagnetic sector, which is not
included here.} 

At order $1/m_Q$, the axial-vector matrix element contains the derivative
correction to the heavy-light current and the time-ordered insertion
of the kinetic operator from the HQET Lagrangian. With the
conventions used below, we write
\begin{align}
    \langle 0|J_\mu^{A}|P(v)\rangle
    &= \langle0|q^\dagger\gamma_\mu\gamma_5 h|P(v)\rangle
    -\frac{1}{2m_Q}
    \langle0|q^\dagger\gamma_\mu\gamma_5\Slash{D}_\perp h|P(v)\rangle
    \nonumber\\
    &\quad -\frac{1}{2m_Q}\int d^4y\,
    \langle0|T\{q^\dagger \gamma_\mu\gamma_5 h(0),\,
    h^\dagger iD_\perp^2 h(y)\}|P(v)\rangle
    +O(1/m_Q^2).
    \label{eq:axial_matrix}
\end{align}
{In Eq.~\eqref{eq:axial_matrix} we keep the ordinary-derivative
  projection $D_\perp\to\partial_\perp$. The gauge-field pieces are
  left to the effective gluonic operator sector. This projection
  affects the long-distance matrix elements but not the need for
  QCD--HQET current matching: the heavy-light axial-vector current is not
  conserved, and its short-distance scale dependence is carried by the
  Wilson coefficients. The matching relation is} 
\begin{align}
    J_\mu^A = \sum_i C_i(\mu_0)\left(J_\mu^{A(i)}
    -\frac{1}{2m_Q}\left(O_\mu^{A(i)}+T_\mu^{A(i)}\right)\right)
    +\mathcal{O}\left(\frac{1}{m_Q^2}\right),
    \label{eq:JA}
\end{align}
where $C_i(\mu_0)$ denote the Wilson coefficients of the HQET
heavy-light axial-vector current at the low-energy scale
$\mu_0$~\cite{Neubert:1993mb}.

The flavor number entering the perturbative running of these Wilson
coefficients is denoted by $n_f^{\rm run}$, to distinguish it from the
single light flavor used in the instanton-induced effective
interaction. Since $\mu_0 \ll m_Q$, we use the renormalization-group
improved expressions~\cite{Neubert:1993za}
\begin{align}
C_1(\mu_0) &= x^{2/\beta_0}
\left[ 1+\frac{\alpha_s(m_Q)-\alpha_s(\mu_0)}{4\pi}S_{\rm hl}
-\frac{4}{3}\frac{\alpha_s(m_Q)}{\pi} \right],\\
C_2(\mu_0)
&= -\frac{2}{3}\,x^{2/\beta_0}\frac{\alpha_s(m_Q)}{\pi},
\quad
x \equiv \frac{\alpha_s(\mu_0)}{\alpha_s(m_Q)},
\end{align}
where
\begin{align}
S_{\rm hl} =
\frac{\gamma_1^{\rm hl}}{2\beta_0}
-\frac{\beta_1\gamma_0^{\rm hl}}{2\beta_0^2},
\quad
\gamma_0^{\rm hl}=-3C_F=-4,
\end{align}
and 
\begin{align}
\gamma_1^{\rm hl}
    = C_F\left[ C_F\left(\frac{5}{2}-16\zeta(2)\right)
    +C_A\left(-\frac{49}{6}+4\zeta(2)\right)
    +\frac{10}{3}T_F n_f^{\rm run} \right] = -\frac{254}{9}-\frac{56\pi^2}{27}+\frac{20}{9}n_f^{\rm run},
\end{align}
with $C_F=\frac{4}{3}$, $C_A=3$, and $T_F=\frac{1}{2}$ for
$\mathrm{SU(3)}_c$. The running 
coupling at two-loop order is given by
\begin{align}
\alpha_s(\mu) = \frac{4\pi}{\beta_0\ln(\mu^2/\Lambda_{\rm QCD}^2)}
\left[
1-\frac{\beta_1}{\beta_0^2}
\frac{\ln\ln(\mu^2/\Lambda_{\rm QCD}^2)}{\ln(\mu^2/\Lambda_{\rm QCD}^2)}
\right],
\end{align}
where
\begin{align}
\beta_0=11-\frac{2}{3}n_f^{\rm run},\quad
\beta_1=102-\frac{38}{3}n_f^{\rm run}.
\end{align}

Each current is defined as
\begin{align}
    J_\mu^{A(1)} &\equiv q^\dagger \gamma_\mu\gamma_5 h,\quad 
    J_\mu^{A(2)} \equiv q^\dagger v_\mu\gamma_5h\\ 
    O_\mu^{A(1)} &\equiv q^\dagger\gamma_\mu\gamma_5\Slash{D}_\perp h,
    \quad O_\mu^{A(2)} \equiv q^\dagger
    v_\mu\gamma_5\Slash{D}_\perp h\\ 
    T_\mu^{A(1)} &\equiv \int d^4y~
    T\Bigl\{J_\mu^{A(1)}(0),\, h^\dagger(y)iD_\perp^2h(y)\Bigl\},\quad
    T_\mu^{A(2)}\equiv\int d^4y~
    T\Bigl\{J_\mu^{A(2)}(0),\, h^\dagger(y)iD_\perp^2h(y)\Bigl\}. 
\end{align}
{The standard dimension-four basis also contains
$O^{A(3,\ldots,6)}$~\cite{Neubert:1993mb}. They first enter at
$O(\alpha_s/m_Q)$, beyond the matching accuracy used here, and are therefore
not included. Below, $F_1$ denotes the retained derivative-current term and
$F_2$ the kinetic insertion in $f_B$; $\lambda_1^{(\partial)}$ refers to the
ordinary-derivative kinetic matrix element.}

The combinations in Eq.~\eqref{eq:JA} can be reduced to single
matrix elements:
\begin{align}
\sum_i C_i(\mu_0)\, \langle0|J_\mu^{A(i)}|P(v)\rangle
    &= [C_1(\mu_0)-C_2(\mu_0)]\, \langle0|J_\mu^{A(1)}|P(v)\rangle\\ 
\sum_i C_i(\mu_0)\, \langle0|O_\mu^{A(i)}|P(v)\rangle
    &= [C_1(\mu_0)+C_2(\mu_0)]\, \langle0|O_\mu^{A(1)}|P(v)\rangle\\ 
\sum_i C_i(\mu_0)\, \langle0|T_\mu^{A(i)}|P(v)\rangle
    &= [C_1(\mu_0)-C_2(\mu_0)]\, \langle0|T_\mu^{A(1)}|P(v)\rangle. 
\end{align}
Applying the Feynman rules, the matrix elements are represented by
\begin{align}
-F_0v_\mu\equiv\langle0| J_\mu^{A(1)}|P(v)\rangle
    &= -{2}G\sqrt{N_c}v_\mu\int_{0}^\infty\frac{dp_4}{\pi}
    \int_0^\infty dp~ |\vec p|^2 F(p)\phi(\vec p)\, \frac{p_4^2-\Lambda M(p)}
    {2\pi^2(p_4^2+\Lambda^2)\left( |\vec p|^2+p_4^2+M^2(p) \right)}\\
-F_1v_\mu\equiv\langle0| O_\mu^{A(1)}|P(v)\rangle
    &= -{2}G\sqrt{N_c}v_\mu\int_0^\infty\frac{dp_4}{\pi}
    \int_0^\infty dp~ |\vec p|^2 F(p)\phi(\vec p)\, \frac{-|\vec p|^2\Lambda}
    {2\pi^2(p_4^2+\Lambda^2)\left(|\vec p|^2+p_4^2+M^2(p)\right)}\\
-F_2v_\mu\equiv\langle0| T_\mu^{A(1)}|P(v)\rangle
    &= {-2}G\sqrt{N_c}v_\mu\int_0^\infty\frac{dp_4}{\pi}\int_0^\infty dp~ 
    |\vec p|^2 F(p)\phi(\vec p)\, \frac{-|\vec p|^2 \left(2p_4^2\Lambda
    +M(p)(p_4^2-\Lambda^2)\right)}{2\pi^2(p_4^2+\Lambda^2)^2
    \left(|\vec p|^2+p_4^2+M^2(p)\right)}.
\end{align}
So, if we define $\mathcal{C}^{\pm}_{12}\equiv C_1\pm C_2$, we get the decay constant $f_{P}=F_{P}/\sqrt{M_{P}}$, given by
\begin{align}
f_{P}=
    \frac{1}{\sqrt{m_{P}}}\left[\mathcal{C}^-_{12}(\mu_0)F_0-\frac{1}
    {2m_Q}\left(\mathcal{C}_{12}^+(\mu_0)F_1
    +\mathcal{C}_{12}^-(\mu_0)F_2\right)\right].
    \label{eq:fP_final}
\end{align}

\section{Kinetic term from the $1/m_Q$ correction}\label{sec:kinetic}

{At order $1/m_Q$, the spin-independent kinetic operator measures
  the residual motion of the heavy quark inside the meson. In the
  present nonlocal model it is weighted by the same vertex profile
  that enters the pole condition, so its ordinary-derivative matrix
  element $\lambda_1^{(\partial)}$ directly probes the momentum
  dependence of the finite-mass kernel. The HQET Lagrangian at this
  order is 
~\cite{Falk:1990pz,Falk:1993dh,Neubert:1993mb,Neubert:1996wg}}
\begin{align}
\mathcal{L}_{1/m_Q}^{\rm HQET} 
    = {+}\frac{i}{2m_Q}h^\dagger\left(D_\perp^2-\frac{1}{2}
    \sigma_{\mu\nu}G_{\mu\nu}\right)h, 
    \label{eq:m_Q_Lag} 
\end{align}
{where $D_{\perp\mu}\equiv D_\mu-v_\mu(v\cdot D)$ and $G_{\mu\nu}$ is the
gluon field strength. The first term defines the full HQET kinetic parameter
$\lambda_1$, while the second represents the chromomagnetic
operator. We evaluate only the ordinary-derivative part of the kinetic term,
$D_\perp^2\to\partial_\perp^2$, and denote the corresponding matrix element
by $\lambda_1^{(\partial)}$. The omitted gauge-field and chromomagnetic
pieces require an effective instanton representation of gluonic operators,
as in the light-quark case~\cite{Diakonov:1995qy,Polyakov:1996kh,Balla:1997hf}.
With this restriction, the pole shift from the $|\vec k|^2$ insertion is}
\begin{align}
    {\delta m_{H}^{(\partial)}}
    =&\frac{\delta S_H^{-1}}{\partial_{i\Lambda}S_H^{-1}}
    =\frac{\frac{1}{2m_Q}G_0^2\Sigma_{\rm kin}(i\Lambda)}{
                             -iG_0^2\Sigma'_P(i\Lambda)}
                             \equiv-\frac{1}{2m_Q}
                             {\lambda_1^{(\partial)}}.
\label{eq:mass_shift}
\end{align}

{Here $\delta S_H^{-1}$ is the first-order change of the inverse
  meson propagator. Expanding the pole condition
  $S_H^{-1}(i\Lambda)=0$ yields the ratio in Eq.~\eqref{eq:mass_shift},
  and the same pole slope that fixes the residue therefore also enters the
  kinetic mass shift.} 

In Eq.~\eqref{eq:mass_shift}, $\lambda_1^{(\partial)}$ (derivative
part) is defined as 
\begin{align}
    \lambda_1^{(\partial)}
    &= \frac{1}{2}\langle H_v|h^\dagger i\partial^2_\perp h|H_v\rangle
    =\frac{G^2}{2}\Sigma_{\rm kin}(i\Lambda)\nonumber\\
    &=i\frac{G^2}{2}\int \frac{d^4k}{(2\pi)^4}~ F^2(k)
    \phi^2(\vec k)\tr_{D}\left[S_q(-k)\Gamma_PS_h(i\Lambda v-k) 
    |\vec k|^2 S_h(i\Lambda v-k)\Gamma_P\right]\nonumber\\
    &= -\frac{G^2}{2}\int_0^\infty\frac{d|\vec k|}{2\pi^2}
    \int_0^\infty \frac{dk_4}{\pi}~ |\vec k|^2F^2(k)\phi^2(\vec k)
    \frac{|\vec k|^2(-2M(k)\Lambda^2+4\Lambda
      k_4^2+2M(k)k_4^2)}{(|\vec
      k|^2+k_4^2+M^2(k))(\Lambda^2+k_4^2)^2}, 
    \label{eq:lambda_1}
\end{align}

{The first line of Eq.~\eqref{eq:lambda_1} defines
$\lambda_1^{(\partial)}$; the remaining lines give its loop representation.
The kinetic insertion supplies the explicit $|\vec k|^2$ weight, making this
quantity particularly sensitive to the high-momentum tail of
$\phi(\vec k)$. It should not be identified with the full $\lambda_1$ until
the gauge-field part of $D_\perp^2$ is included.}

\section{Isgur-Wise function}\label{sec:iw}

Heavy-quark symmetry implies that all heavy-to-heavy semileptonic form
factors are governed, at leading order in $1/m_Q$, by a single
universal function of the recoil parameter
$w=v\cdot v'$~\cite{Isgur:1989vq,Isgur:1990yhj,Ligeti:1993hw,Caprini:1995wq}.
A nontrivial check of any heavy-light bound-state model is therefore
whether it reproduces the normalization $\xi(1)=1$ at zero recoil and
yields a reasonable slope
$\rho_{\rm IW}^2=-\left.d\xi(w)/dw\right|_{w=1}$.
In the present work we do not introduce separate finite-mass
heavy-light vertex form factors for different heavy flavors. Instead,
the Isgur-Wise function is evaluated directly in the static
heavy-quark limit with the same heavy-light vertex form factor
$\phi(\vec p)$ that appears in the heavy-light two-point function.

\begin{figure}[htp!]
     \centering
    \includegraphics[width=6cm]{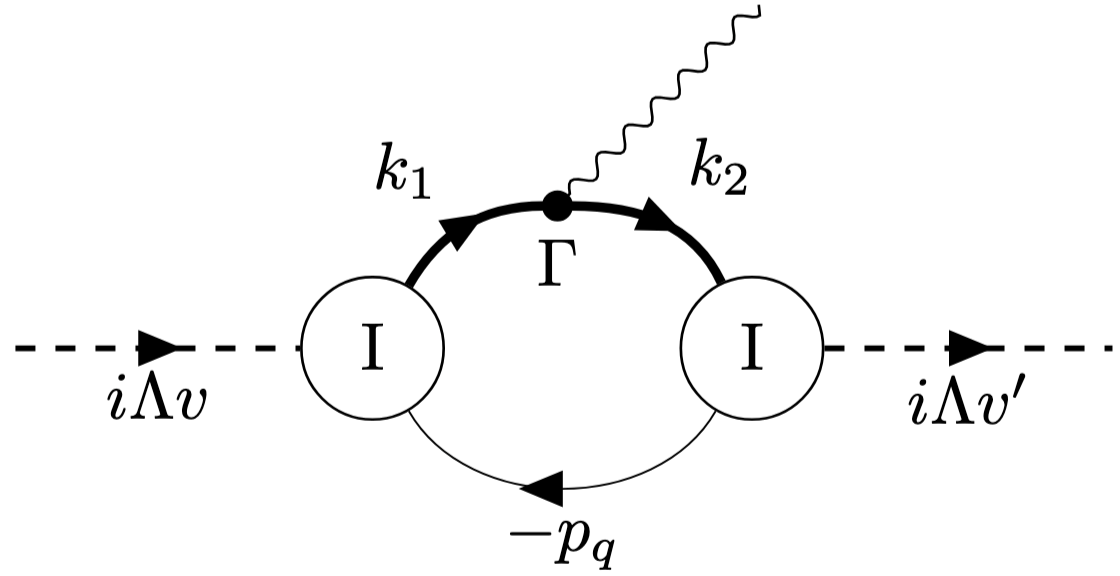}
    \caption{Feynman diagram for the heavy-to-heavy form factor. The
    blob $I$ is the nonlocal heavy-light vertex induced by the
    instanton vacuum.}
    \label{fig:Isgur-Wise}
\end{figure}

The heavy-to-heavy transition is represented by the static current
$-ih_{v'}^\dagger\Gamma_\mu h_v$, where $v$ and $v'$ denote the
velocities of the initial and final static heavy fields. The corresponding Feynman
diagram is shown in Fig.~\ref{fig:Isgur-Wise}.

For the pseudoscalar case, the static-limit vertex representation gives
\begin{align}
    -i\langle H(v')|h_{v'}^\dagger\Gamma_\mu h_v|H(v)\rangle
    =-iG^2\int \frac{d^4p_q}{(2\pi)^4}~
    F^2(p_q)\phi^2(\vec p_q) \,\tr_D\left[\Gamma_P S_q(-p_q)\Gamma_P
    S_h(i\Lambda v'-p_q)\Gamma_\mu
    S_h(i\Lambda v-p_q)\right]. 
    \label{eq:IW_mat}
\end{align}
Here the two heavy propagators are the HQET propagators of
Eq.~\eqref{eq:propagators} with velocities $v'$ and $v$, respectively. The
same coupling $G$, residual mass $\Lambda$, and heavy-light vertex form factor
$\phi(\vec p)$ are used at both heavy-light vertices, because the
quantity extracted here is the universal static-limit function rather
than a finite-mass flavor-changing form factor.

To calculate the correlation function, we must define the velocities
$v$ and $v'$. In Euclidean convention, we use
$v^0_M=v_{E4}$ and $v^i_M=iv_{Ei}$, then the velocity satisfies
$v^2_M=v_E^2=1$ and
$\frac{1+\Slash{v}_M}{2}=\frac{1+\Slash{v}_E}{2}$. If we set the
Euclidean velocity $v=(\vec{0},1)$ as an initial state, then the final
state in the COM frame has $v'$ as
\begin{align}
    v'=(0,0,\sqrt{1-w^2},w), 
\end{align}
where $v'^2=1$ and $w=v\cdot v'$.
With the Euclidean normalization $v^2=v'^2=1$, one has $w=\cos\theta$
and therefore $-1\le w \le 1$ in purely Euclidean kinematics.
Physical recoil corresponds to the Minkowski region $w\ge 1$.
In this nonlocal model, however, a direct continuation of the Euclidean
kinematics to $w>1$ can lead to unstable or unphysical behavior because
the heavy-line contribution and the nonlocal form factors become
complex. For this reason we evaluate the loop integrals only for
$-1\le w\le 1$ in Euclidean kinematics and extract the recoil
dependence from the behavior around zero recoil.

Hence, we can rewrite the correlation function~\eqref{eq:IW_mat} as
\begin{align}
     -i\langle H(v')|h_{v'}^\dagger\Gamma_\mu h_v|H(v)\rangle
     =-iG^2\tr_D\left[\gamma_5\frac{1+\Slash{v}'}{2}\Gamma_\mu
     \frac{1+\Slash{v}}{2}\gamma_5\mathcal{M}\right],
\end{align}
where $\mathcal{M}$ is defined as
\begin{align}
    \mathcal{M}
    \equiv \int \frac{d^4p_q}{(2\pi)^4}~
    \frac{-\Slash{p}_q+iM(p_q)}{p_q^2+M^2(p_q)}
    \frac{F^2(p_q)\phi^2(\vec p_q)}
    {(i\Lambda-v\cdot p_q)(i\Lambda-v'\cdot p_q)}.
\end{align}
Having computed the trace over spin space, we obtain
\begin{align}
    -i\langle H(v')|h_{v'}^\dagger\Gamma_\mu h_v|H(v)\rangle
    &=-iG^2\int \frac{d^4p}{(2\pi)^4}~
    \frac{-(v'\cdot v-1)p_{\mu}+[v_\mu (v'\cdot p)+v'_\mu (v\cdot
      p)]+iM(p)(v_\mu+v'_\mu)} 
    {p^2+M^2(p)}\cr
    &\quad\times \frac{F^2(p)\phi^2(\vec p)}
    {(i\Lambda-v\cdot p)(i\Lambda-v'\cdot p)}.
    \label{eq:IW_mat_int}
\end{align}
Guided by the standard HQET decomposition of the
pseudoscalar heavy-to-heavy matrix element, we project the loop result
onto the coefficient of $(v+v')_\mu$. Although the separable vertex
$\phi(\vec p)$ is implemented in the meson rest frame and does not
make full covariance manifest, the leading heavy-quark-symmetry
structure can be written as
$F(v,v') = f_+(w)(v+v') + f_-(w)(v-v')$~\cite{Neubert:1993mb,Neubert:1996wg}.
Therefore, Eq.~\eqref{eq:IW_mat_int} is rewritten in this standard HQET basis.
To obtain this form, the momentum $p$ can be represented in the basis
of $v$ and $v'$,
\begin{align}
    p_{\mu} = a(v+v')_\mu+b(v-v')_\mu+p_{\mu}^\perp, \label{eq:HQET_mom}
\end{align}
where the perpendicular momentum $p^\perp$ satisfies
$v\cdot p^\perp=0$ and $v'\cdot p^\perp=0$. Multiplying by
$(v+v')$ and $(v-v')$, respectively, gives
\begin{align}
    a=\frac{(v+v')\cdot p}{2(1+v\cdot v')},\quad
    b=\frac{(v-v')\cdot p}{2(1-v\cdot v')}.
\end{align}
Then Eq.~\eqref{eq:IW_mat_int} is reduced to the following simple relation: 
\begin{align}
    -i\langle H(v')&|h_{v'}^\dagger\gamma_\mu h_v|H(v)\rangle\cr
    =&-iG^2\int\frac{d^4p}{(2\pi)^4}
    \frac{F^2(p)\phi^2(\vec p)}
    {(p^2+M^2(p))(i\Lambda-v\cdot p)(i\Lambda-v'\cdot p)}
    \left[iM(p)+\frac{(v+v')\cdot p}{1+v\cdot v'}\right]
    (v+v')_\mu\cr
    \equiv&\xi(w)(v+v')_\mu,
    \label{eq:IW_mat_rel}
\end{align}
where $\xi(w)$ denotes the universal Isgur-Wise function in the present
separable approximation. The result is generalized as
\begin{align}
    \langle H'(v')|h_{v'}^\dagger\Gamma_\mu h_v|H(v)\rangle
    = i\xi(w)\Tr\left[\Gamma'\frac{1+\Slash{v}'}{2}\Gamma_\mu
    \frac{1+\Slash{v}}{2}\Gamma\right].
\end{align}
The decomposition in Eq.~\eqref{eq:HQET_mom} defines the Isgur-Wise function
as the coefficient of $(v+v')_\mu$ in the heavy-quark limit.
At zero recoil, $v=v'$, and the current becomes the 
heavy-flavor symmetry generator. In this limit, Eq.~\eqref{eq:IW_mat_rel}
reduces, with the same Euclidean sign convention as in
Eq.~\eqref{eq:inv_meson_prop}, to
\begin{align}
    \xi(1)=\frac{iG^2}{2}\Sigma'_P(i\Lambda).
\end{align}
The residue normalization $iG^2\Sigma'_P(i\Lambda)=2$ then gives
$\xi(1)=1$. This is a nontrivial check for our framework because the
same residue-normalized vertex coupling is used in both the two-point 
and three-point functions.

For nonzero recoil, the $w$ dependence of $\xi(w)$ is determined by the
heavy-light vertex form factor together with the nonlocal structure of
the current insertion. In Euclidean kinematics, one may write
$w_E=\cos\theta=1-\theta^2/2+\cdots$, whereas the physical Minkowski
recoil variable is $w_M=\cosh\eta=1+\eta^2/2+\cdots$. Thus the
Euclidean and Minkowski regions approach the common zero-recoil point
$w=1$ from opposite sides. {If the loop amplitude is analytic in a
  neighborhood of $w=1$, the zero-recoil derivative is fixed
  unambiguously. We use the standard Isgur-Wise}
slope~\cite{Isgur:1990jf,Caprini:1994np,Neubert:1992kk} 
\begin{align}
    \rho_{\rm IW}^2
    \equiv -\left.\frac{d\xi(w)}{dw}\right|_{w=1}.
\end{align}
Equivalently, near zero recoil, the same analytic function is expanded as
\begin{align}
    \xi(w)=1-\rho_{\rm IW}^2(w-1)+O\!\left((w-1)^2\right).
\end{align}
In the Euclidean calculation, this expansion is evaluated for $w_E<1$,
so that
\begin{align}
    \xi_E(w_E)
    = 1-\rho_{\rm IW}^2(w_E-1)+O\!\left((w_E-1)^2\right)
    = 1+\rho_{\rm IW}^2(1-w_E)+O\!\left((1-w_E)^2\right).
\end{align}
Thus, the Euclidean derivative gives
\begin{align}
    \rho_{\rm IW}^2
    = -\left.\frac{d\xi_E(w_E)}{dw_E}\right|_{w_E=1^-}
\end{align}
This identification concerns the derivative with respect to $w$, not
the coefficient of an expansion in the Euclidean angle $\theta$ or the
Minkowski rapidity $\eta$. The derivative is evaluated from Euclidean
loop integrals in the interval $-1\le w_E\le 1$.

\section{Numerical Results}\label{sec:numerics}

We present a representative numerical solution of this framework. The
instanton parameters are kept fixed at $M_q=345\,\mathrm{MeV}$,
$\bar\rho^{-1}=600\,\mathrm{MeV}$, and $n=(200\,\mathrm{MeV})^4$,
with the standard dynamical-mass input discussed above~\cite{Diakonov:1985eg,Diakonov:2002fq}.
For the heavy-quark sector, $m_b^{\rm eff}=5.04\,\mathrm{GeV}$ is used in the
leading mass relation, while $m_b=4.695\,\mathrm{GeV}$ is used in the
short-distance matching and in the explicit $1/m_b$ terms. The finite-mass
heavy-line kernel is represented by the same heavy-light vertex form factor
that enters the pole condition, the decay constant, the kinetic correction,
and the Isgur-Wise matrix element. For the numerical example, we use
\begin{align}
    \phi(\vec k)=\frac{1}{\left(1+|\vec k|^2/b^2\right)^a}
    .
    \label{eq:phi_numerical}
\end{align}

{
Equation~\eqref{eq:phi_numerical} is our central power-law (PL) choice. The
separable ansatz does not determine the detailed momentum dependence of the
finite-mass vertex, however, so using only one shape would leave that choice
untested. We therefore refit two alternatives. The first is the rational (R)
form
\begin{align}
    \phi_{\rm R}(\vec k)
    &=\frac{1}{1+\left(|\vec k|/b\right)^a},
    \label{eq:phi_R}
\end{align}
and, for a qualitatively different falloff, we use the Bessel-inspired (B)
form
\begin{align}
    \phi_{\rm B}(\vec k)
    &=\left[\mathcal B(z_B)\right]^a,
    \qquad z_B\equiv\frac{|\vec k|}{2b},
    \label{eq:phi_B}\\
    \mathcal B(z)
    &\equiv 2z\left[I_0(z)K_1(z)-I_1(z)K_0(z)\right]
    -2I_1(z)K_1(z),
    \qquad \mathcal B(0)=1.
    \label{eq:Bessel_profile_kernel}
\end{align}
Only the shape of $\mathcal B(z)$ is borrowed from the modified-Bessel
structure of the instanton zero mode. Its parameters $a$ and $b$ are fitted
anew. No extra light-quark factor $F(k)$ is introduced, so $\phi_{\rm B}$ is
simply another finite-mass heavy-light vertex profile. All three choices obey
$\phi(0)=1$.
}

{Each profile is fitted independently. For PL, $b$ fixes the
  momentum scale and $a$ the power of the falloff. The pole condition
  used in the fit is} 
\begin{align}
    G_0^2\, \Sigma_P(i\Lambda) = 1
    \label{eq:pole_G0_numerical}
\end{align}
{and the residue-normalized coupling}
\begin{align}
    G^2 = \frac{2}{i\Sigma'_P(i\Lambda)}.
    \label{eq:G_rescaled_numerical}
\end{align}
{Equivalently,
$\mathcal{R}_H=i\Sigma'_P(i\Lambda)/\Sigma_P(i\Lambda)$, and we impose
$\mathcal{R}_H=2\,\mathrm{GeV}^{-1}$. Together with
$f_B^{\rm LO}=180.0\,\mathrm{MeV}$ and
$\bar m_B=5.313\,\mathrm{GeV}$, these conditions determine the profile
parameters. Fits are restricted to $0.25\leq a\leq20$ and
$0.015\leq b\leq3\,\mathrm{GeV}$.}

\begin{figure}[htp!]
    \centering
    \includegraphics[width=0.62\textwidth]{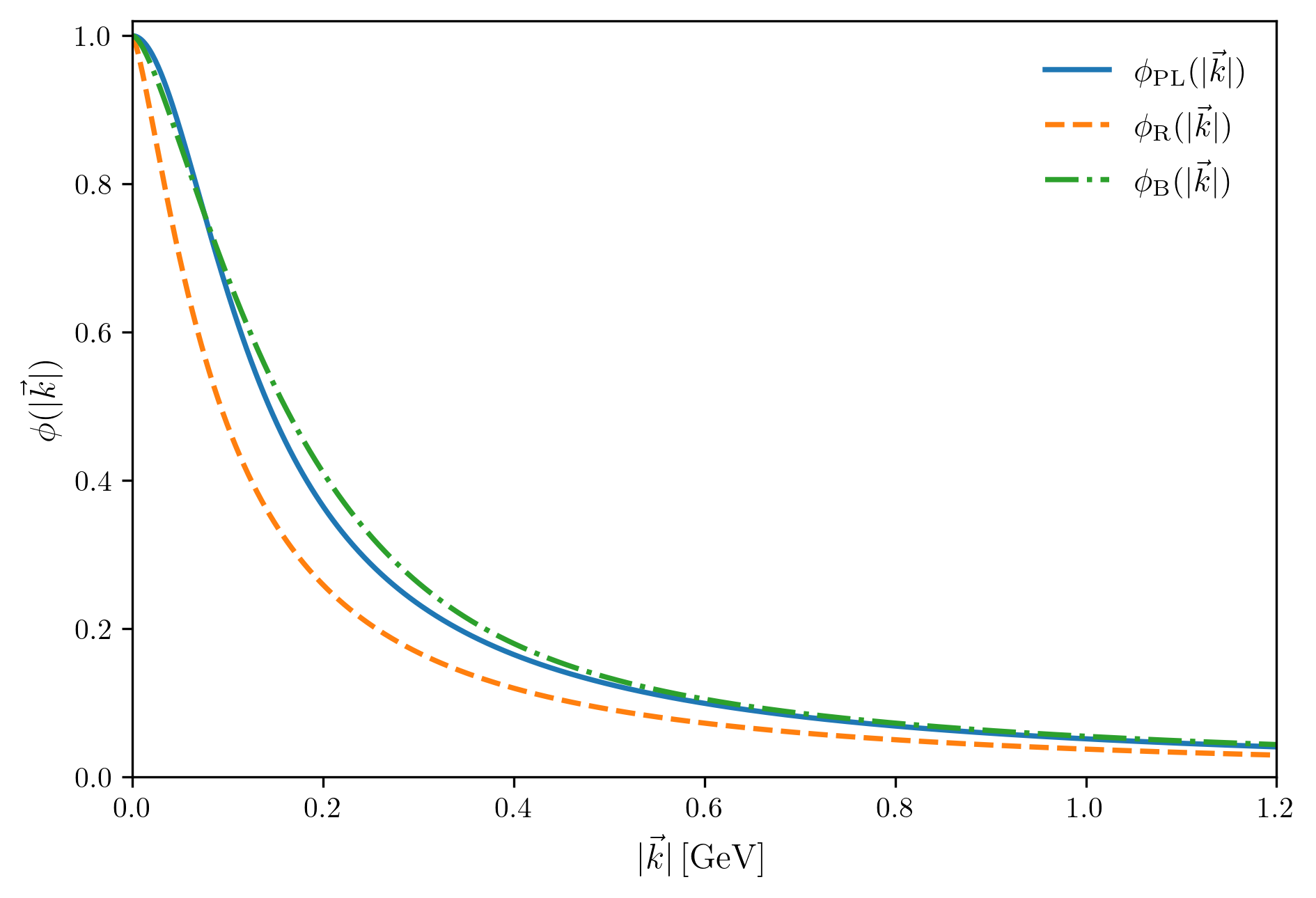}
    \caption{Fitted PL, R, and B heavy-light vertex profiles. The three
    forms satisfy the same residue, $f_B^{\rm LO}$, and spin-averaged mass
    conditions.}
    \label{fig:profile_robustness}
\end{figure}

Table~\ref{tab:profile_robustness} gives the fitted $a$ and $b$ values;
Fig.~\ref{fig:profile_robustness} shows the corresponding profiles.

\begin{table}[htp!]
\centering
\caption{Fitted parameters for the three finite-mass heavy-light vertex
profiles. Each profile is calibrated to the same residue, $f_B^{\rm LO}$,
and spin-averaged mass conditions.}
\label{tab:profile_robustness}
\begin{tabular}{c|cc}
\hline\hline
Profile & $a$ & $b$ [GeV]\\
\hline
PL & 0.6560 & 0.1048\\
R  & 1.3631 & 0.0926\\
B  & 0.4103 & 0.0517\\
\hline\hline
\end{tabular}
\end{table}

Having refitted, we find that the three shapes yield very similar
observables. Relative to PL, the largest shifts are $1.3\%$ in
$\Lambda$, $2.9\%$ in $-\lambda_1^{(\partial)}$, $1.3\%$ in $\rho_{\rm
  IW}^2$, and $4.0\%$ in the retained partial decay correction. We use
these shifts as the functional-form spread of the vertex ansatz. 

Table~\ref{tab:instanton_sensitivity} summarizes the sensitivity to the instanton inputs, with the PL functional form retained. Our central choice is close to the conventional Diakonov--Petrov
set ~\cite{Diakonov:1985eg,Diakonov:2002fq}. Note that the set of 
Ref.~\cite{Liu:2024rdm} uses $M_q=398\,\mathrm{MeV}$ and
$\bar\rho=0.313\,\mathrm{fm}$. In each case $a$ and $b$ are refitted. Values
from the two parameterizations are not mixed.

\begin{table*}[htp!]
\centering
\caption{Sensitivity to two representative instanton-vacuum parameter
sets used in the literature.  The PL profile is independently refitted
for each row to $\mathcal R_H=2\,\mathrm{GeV}^{-1}$,
$f_B^{\rm LO}=180.0\,\mathrm{MeV}$, and
$\bar m_B=5.313\,\mathrm{GeV}$.}
\label{tab:instanton_sensitivity}
\resizebox{\textwidth}{!}{%
\begin{tabular}{c|cc|cccccc}
\hline\hline
Set & $M_q$ [MeV] & $\bar\rho$ [fm] & $a$ & $b$ [GeV]
& $\Lambda$ [MeV] & $-\lambda_1^{(\partial)}$ [GeV$^2$]
& $\rho_{\rm IW}^2$
& $\delta f_B^{(1/m_Q,\mathrm{partial})}$ [MeV]\\
\hline
Present~\cite{Diakonov:1985eg,Diakonov:2002fq}
& 345 & 0.329 & 0.6560 & 0.1048 & 190.7 & 0.7730 & 1.1529 & 121.9\\
Ref.~\cite{Liu:2024rdm}
& 398 & 0.313 & 0.7446 & 0.1835 & 185.6 & 0.8211 & 1.2966 & 120.8\\
\hline\hline
\end{tabular}}
\end{table*}

The partial decay correction hardly moves between the two sets: its change
is $0.9\%$. By comparison, $\Lambda$ changes by $2.7\%$,
$-\lambda_1^{(\partial)}$ by $6.2\%$, and the recoil slope by $12.5\%$.
Thus $\rho_{\rm IW}^2$ carries the largest input dependence in this test.
We do not scan the instanton density independently. At fixed $M_q$ and
$\bar\rho$, the explicit $n$ dependence of $g^2=M_q\Delta M_Q/(16n)$
cancels through $\Delta M_Q\propto n$, while $G_0$ is fixed by the pole
condition.

{The chromomagnetic sector is not included in this calculation, so
  the mass constraint is applied to the spin average}
\begin{align}
    \bar m_B = \frac{m_B+3m_{B^*}}{4}.
\end{align}
{This removes the leading hyperfine splitting. With the
  heavy-quark inputs specified above, the fitted mass relation
  becomes} 
\begin{align}
    \bar m_B = m_b^{\rm eff}+\Lambda
    {-\frac{\lambda_1^{(\partial)}}{2m_b}},
    \label{eq:spin_averaged_mass}
\end{align}
{The same fit also imposes the residue condition and
$f_B^{\rm LO}=180.0\,\mathrm{MeV}$. The resulting quantities are collected
in Table~\ref{tab:numerical_results}.}
\begin{table*}[htp!]
\centering
\caption{Representative numerical results.  The quoted reference values
are used only for orientation.  The values taken from the literature may
correspond to different mass schemes and conventions.  In particular,
the entries in the $\Lambda$ column should be interpreted as the usual
HQET parameter $\Lambda$ in those references.  The range quoted for
$-\lambda_1$ is scheme-dependent and is taken from representative 
HQE/HQET analyses~\cite{Jeong:1998gj,Chernyak:1996bq,Gambino:2013rza},
while the Isgur-Wise slope bound is the QCD sum-rule lower 
bound~\cite{Uraltsev:2000ce}.  The Belle-II value is the
experimentally fitted slope parameter at zero
recoil~\cite{Belle-II:2025rna}.} 
\label{tab:numerical_results}
\begin{tabular}{c|cccccccc}
\hline\hline
 Profile & $f_B^{\rm LO}$ [MeV] & {$\delta
    f_B^{(1/m_Q,\mathrm{partial})}$}
    [MeV]&
           {$f_B^{\rm  partial}$}
           [MeV] 
 & {$f_B^{\rm ref}$ [MeV]} & $\Lambda$ [MeV] & $m_b^{\rm eff}$ [GeV]
 &  $-\lambda_1^{(\partial)}$ [GeV$^2$] & $\rho_{\rm IW}^2$\\
\hline
{ PL} & {180.0} & {$+121.9$}
&  {301.9} & {--}
&  {190.7} & 5.04
&  {0.773}
&  {1.153}\\
{ R} & {180.0} & {$+119.8$}
&  {299.8} & {--}
&  {189.6} & {5.04}
&  {0.783}
&  {1.145}\\
{ B} & {180.0} & {$+126.8$}
&  {306.8} & {--}
&  {193.1} & {5.04}
&  {0.751}
&  {1.168}\\
\hline
PDG~\cite{ParticleDataGroup:2024cfk} & & &  & {190.0} & - & - & - & - \\
Ref.~\cite{Jeong:1998gj}   & & & & - & 314-588 & 4.70-4.90   & 0.26-0.97    & - \\
Ref.~\cite{Chernyak:1996bq} & & & & - & 240-320 & 5.02         & 0.11-0.17    & - \\
Ref.~\cite{Gambino:2013rza} & & & & - & -       & 4.518-4.564 & 0.336-0.492  & - \\
Ref.~\cite{Uraltsev:2000ce} & & & & - & -       & -           & -            & $>0.75$ \\
Belle-II~\cite{Belle-II:2025rna}
                            & & & & - & -       & -           & -            & $1.09\pm0.06$ \\
\hline\hline
\end{tabular}
\end{table*}

{Two entries in Table~\ref{tab:numerical_results} should be distinguished
from the phenomenological reference columns. The quantity
$f_B^{\rm partial}$ contains only the retained derivative-current and kinetic
insertions, so the PDG value is shown separately as $f_B^{\rm ref}$.
Likewise, the quoted literature values of $-\lambda_1$ are for the full
covariant HQET kinetic matrix element, not for our
$-\lambda_1^{(\partial)}$.

For PL, the pole condition gives $\Lambda=190.7\,\mathrm{MeV}$. Together
with $\lambda_1^{(\partial)}=-0.773\,\mathrm{GeV}^2$, the retained kinetic
piece contributes}
\begin{align}
    {-\frac{\lambda_1^{(\partial)}}{2m_b}}
    \simeq \;
    {82.3\,\mathrm{MeV}},
\end{align}
{which is approximately $43\%$ of $\Lambda$.
The retained partial decay-constant correction,
$\delta f_B^{(1/m_Q,\mathrm{partial})}=+121.9\,\mathrm{MeV}$,
amounts to approximately $68\%$ of the leading-order value
$f_B^{\rm LO}=180.0\,\mathrm{MeV}$.
These results show that the retained derivative sector is
numerically important. Neither quantity, however, represents
the complete correction at order $1/m_b$.
Their magnitudes alone therefore do not establish a breakdown
of the heavy-quark expansion; a quantitative assessment requires
the omitted contributions at this order and an estimate of
higher-order effects.

After independent refits to the same conditions, the PL, R, and B
profiles yield partial decay-constant corrections between
$119.8$ and $126.8\,\mathrm{MeV}$, with a maximum deviation of
$4.0\%$ from the PL result.
Thus, the sizable partial correction persists across the three
tested functional forms, with limited profile dependence under
the adopted fitting conditions.}

For the recoil dependence, the loop calculation fixes the zero-recoil
slope. We therefore quote \(\rho_{\rm IW}^2\) as the primary result.
When an illustrative recoil-shape representation is needed, we use the
slope-preserving parametrization
\begin{equation}
    \xi_{\rm fit}(w) = \left(\frac{2}{1+w}\right)^{2\rho_{\rm IW}^2}.
\end{equation}
With \(\rho_{\rm IW}^2=\,{1.153}\), this
parametrization is normalized at zero recoil by construction.
It should be understood only as a
reconstruction from the fitted zero-recoil slope, not as an
independent calculation of the full recoil dependence. The slope
value satisfies the QCD sum-rule lower bound~\cite{Uraltsev:2000ce}
and is close to the Belle-II slope quoted in
Table~\ref{tab:numerical_results}.

\section{Summary and outlook}\label{sec:summary}
We have formulated a nonlocal instanton-based description of
heavy-light mesons in which the static Wilson-line form factor
$F_Q^{(\infty)}(\vec q)$ is clearly separated from the effective
finite-mass heavy-line vertex used in the calculation. The separable
effective kernel represents the finite-mass vertex $\mathcal{F}(\vec
p,\vec l;m_Q)\simeq\phi(\vec p)\phi(\vec l)/N_\phi^2$, where
$\phi(\vec p)$ is the dimensionless heavy-light vertex profile and
$N_\phi$ is its dimensionless overall normalization factor. With this
representation, the heavy-light current is written as a single
four-momentum integral over the relative momentum, and the
Hubbard-Stratonovich field is introduced as an auxiliary field
depending only on the total momentum. At the same time, the nonlocal
relative-momentum structure is kept in the current.

{Only the combination $G_0=g/N_\phi$ enters the separable
interaction, so $N_\phi$ need not be fixed separately. The pseudoscalar
pole condition $G_0^2\Sigma_P(i\Lambda)=1$ fixes $G_0$, while the residue
factor ${\cal R}_H=iG_0^2\Sigma'_P(i\Lambda)$ supplies the dimensional
rescaling from the auxiliary bosonized field to the canonical HQET meson
field. We impose ${\cal R}_H=2\,\mathrm{GeV}^{-1}$ as this normalization
condition and define $G=\sqrt{2/{\cal R}_H}\,G_0$.} The same residue-normalized
vertex $G F(p)\phi(\vec p)\Gamma_i$ enters the decay constant, the
kinetic $1/m_Q$ correction, and the heavy-to-heavy matrix element. The
framework, therefore, treats the two-point function and the matrix
elements within one common nonlocal structure while keeping the
pre-residue coupling $G_0$ distinct from the physical coupling $G$.

For the representative vertex form factor in
Eq.~\eqref{eq:phi_numerical},
we obtained {$f_B^{\mathrm{LO}}=180.0\,\mathrm{MeV}$,
$\delta f_B^{(1/m_Q,\mathrm{partial})}=+121.9\,\mathrm{MeV}$,}
{$f_B^{\rm partial}=301.9\,\mathrm{MeV}$},
$\Lambda=\,{190.7}\,\mathrm{MeV}$,
$m_b^{\rm eff}=5.04\,\mathrm{GeV}$ {(fixed)} for
$\bar m_B=\,{5.313}\,\mathrm{GeV}$,
$\lambda_1^{(\partial)}=-\,{0.773}
\,\mathrm{GeV}^2$, and
$\rho_{\rm IW}^2=\,{1.153}$.
{The leading decay constant and the spin-averaged mass are used
in the fit, while the remaining quantities follow from the fitted profile.
The R and B profiles give only a few-percent spread around the central PL
result: the largest changes are $1.3\%$ in $\Lambda$, $2.9\%$ in
$-\lambda_1^{(\partial)}$, $1.3\%$ in $\rho_{\rm IW}^2$, and $4.0\%$ in
the retained partial decay-constant correction. Changing to the alternative
instanton-vacuum parameter set has a larger effect on the Isgur-Wise slope,
which shifts by $12.5\%$, while the partial correction to $f_B$ changes by
less than $1\%$.}

The kinetic contribution to the residual mass is about $43\%$ of the
leading residual contribution $\Lambda$, and the corresponding
{partial $1/m_Q$ correction} to the
decay constant is numerically sizable. {This partial result
should not be interpreted as a complete $O(1/m_b)$ prediction; rather,
it shows that the retained spin-independent subleading sector is a
sensitive probe of the finite-mass heavy-light vertex.} The kinetic
insertion is treated explicitly here because it is directly tied to the
momentum dependence of the finite-mass heavy-light vertex.
{The remaining gauge-field and spin-dependent contributions can
be incorporated through the corresponding effective gluonic operators and
matching conditions within the same instanton-based framework.} This
extension is under investigation.

\section*{Acknowledgment}
The present work was supported by the National Research Foundation of
Korea (NRF) grant funded by the Korea government under Grant
No. RS-2025-02634319 (KHH and YC) and RS-2025-00513982 (HChK and NR).

\appendix
\section{Wilson-line kernel}\label{app:A}

Here we specify the details of the Wilson-line calculation entering
Eq.~\eqref{eq:S_int}. In the heavy-quark static limit, the free heavy-quark
propagator can be written in terms of the operator $\theta$ as
\begin{align}
    S_0(x,y) &= \theta(x_4-y_4) \, \delta^{(3)}(\vec x-\vec y),\\
    S_0^{-1}(x,y) &= \delta^{(4)}(x-y) \, \partial_{y_4},
\end{align}
where $\theta(x_4-y_4)$ denotes the Heaviside function. The propagator in
the presence of a single instanton field $A_I$ is
\begin{align}
    S_A(x,y)
    &=\theta(x_4-y_4) \, \delta^{(3)}(\vec x-\vec y) \,
    W(\vec x-\vec z;\, x_4,\, y_4),\\
    W(\vec x-\vec z;\, x_4,\, y_4)
    &=\mathcal{P}\exp\!\left[i\int_{y_4}^{x_4}d\tau\,\,
    A_{I4}(\vec x-\vec z,\, \tau-z_4)\right],
\end{align}
where the instanton gauge field in singular gauge is given by
\begin{align}
    A_{I\mu}(x) =
    \frac{\bar{\eta}^a_{\mu\nu}\tau^a x_\nu\rho^2}{x^2(x^2+\rho^2)},
\end{align}
and only the temporal component $A_{I4}$ enters the static Wilson line.
Here $\bar{\eta}^a_{\mu\nu}$ denotes the 't Hooft symbol.

In momentum space, the heavy-quark contribution reads
\begin{align}
    &\int d^4x\, d^4y~ h^{\dagger}(y)
    \tr_c\!\left[\langle y|\theta^{-1}(w-\theta)\theta^{-1}|x\rangle
    \right]h(x)\nonumber\\
    &\qquad = \int\frac{d^4p_{2}}{(2\pi)^{4}}
    \frac{d^4p_{1}}{(2\pi)^{4}} h^{\dagger}(p_{2})
    \int d^{3}x~ e^{-i(\vec p_{2}-\vec p_{1})\cdot\vec x}\,
    \tr_c \left[\int dx_{4}\,dy_{4}\, e^{-i\omega_{p_2}y_{4}+i\omega_{p_1}x_{4}}
    \langle y_4|\theta^{-1}(w-\theta)\theta^{-1}|x_4\rangle\right] h(p_{1}).
\end{align}

To evaluate the Wilson-line operator, we use
\begin{align}
    \langle t_1|\theta|t_2\rangle
    = \theta(t_1-t_2), \quad
    \langle t_1|\theta^{-1}|t_2\rangle
    = -\partial_{t_1}\delta(t_1-t_2).
\end{align}
Inserting the completeness relation
$\int d\tau~|\tau\rangle\langle\tau|=1$, one obtains
\begin{align}
    &\int dx_{4}\,dy_{4}~ e^{-i\omega_{p_2}y_{4}+i\omega_{p_1}x_{4}}
    \langle y_4|\theta^{-1}
    \left(\int d\tau_{2}~|\tau_{2}\rangle\langle\tau_{2}|\right)
    (w-\theta)\left(\int d\tau_{1}~|\tau_{1}\rangle\langle\tau_{1}|\right)
    \theta^{-1}|x_4\rangle \nonumber\\
    &\qquad = -\omega_{p_1}\omega_{p_2}\,
    e^{-i(p_2-p_1)\cdot z}\int_{-\infty}^{\infty} dT_1
    \int_{-\infty}^{\infty} dT_2~ e^{-i\omega_{p_2}T_2+i\omega_{p_1}T_1}
    \,\theta(T_2-T_1) 
    \left\{ \mathcal{P}\exp\!\left[ i\int_{T_1}^{T_2}dr_4\,
    \frac{\bar{\eta}^{a}_{4\nu}\tau^a r_\nu\rho^2}
    {r^2(r^2+\rho^2)} \right]-1 \right\},
    \label{eq:Pexp}
\end{align}
where
$r_\mu=(\vec x-\vec z,\, \tau-z_4)$ and
$T_{1,2}=\tau_{1,2}-z_4$.

After taking the color trace, Eq.~\eqref{eq:Pexp} becomes
\begin{align}
    -2\omega_{p_1}\omega_{p_2}\,
    e^{-i(p_2-p_1)\cdot z}
    \int_{-\infty}^{\infty} dT_1 \int_{-\infty}^{\infty} dT_2~
    e^{-i\omega_{p_2}T_2+i\omega_{p_1}T_1}\,
    \theta(T_2-T_1)\, \sin^2\!\Big[
    \frac{1}{2}\left(\Omega(T_2)-\Omega(T_1)\Big) \right],
    \label{eq:A9}
\end{align}
with
\begin{align}
    \Omega(T)
    \equiv\tan^{-1}\!\left(\frac{T}{r}\right)
    -\frac{r}{\sqrt{r^2+\rho^2}}\tan^{-1}\!\left(
    \frac{T}{\sqrt{r^2+\rho^2}}\right).
\end{align}

With the HQET momentum decomposition
$P_Q^\mu=m_Qv^\mu+k^\mu$ ($v^2=1$),
where $k^\mu$ denotes the residual momentum and the residual
energies satisfy $\omega_{1,2}\sim v\!\cdot\!k\ll m_Q$.
Since the heavy quark remains close to its mass shell,
\begin{align}
    P_Q^2
    = m_Q^2+2m_Qv\!\cdot\!k+k^2 \simeq m_Q^2,
\end{align}
the Wilson-line kernel may be evaluated at
$\omega_{1,2}=0$ to leading order in the HQET expansion.

In this limit the characteristic time scale is
$T\sim 1/\omega$, which is parametrically large. Since the instanton
form factor suppresses hard contributions, the dominant support of the
time integrals comes from the large-$|T|$ region. Therefore,
\begin{align}
    \lim_{T\rightarrow\pm\infty}\Omega(T)
    = \pm\frac{\pi}{2}
    \left( 1-\frac{r}{\sqrt{r^2+\rho^2}} \right)
    \equiv \Omega_\infty(r)\,\mathrm{sgn}(T),
\end{align}
so that
\begin{align}
    \Omega(T_2)-\Omega(T_1)
    \simeq \Omega_\infty(r)
    \Big[ \mathrm{sgn}(T_2)-\mathrm{sgn}(T_1) \Big].
\end{align}

The difference vanishes when $T_1$ and $T_2$ have the same sign, so
the only nonvanishing contribution comes from the region
$T_1<0<T_2$. Consequently,
\begin{align}
    &-2 \omega_{p_1}\omega_{p_2}
    \int_{-\infty}^{0} dT_1 \int_{0}^{\infty} dT_2~
    e^{-i\omega_{p_2}T_2+i\omega_{p_1}T_1}
    \int d^3r~ e^{-i(\vec p_1-\vec p_2)\cdot\vec r}\,
    \sin^2\Big[ \Omega_\infty(r) \Big] \nonumber\\
     &\qquad = 8\pi \int_0^\infty dr~ r^2
    j_0(|\vec p_1-\vec p_2|r)\, \sin^2\!\left(
    \frac{\pi}{2}-\frac{\pi r}{2\sqrt{r^2+\rho^2}}\right)
    = 8\pi \int_0^\infty dr~ r^2
    j_0(|\vec p_1-\vec p_2|r)\, \cos^2\!\left(\frac{\pi r}{2\sqrt{r^2+\rho^2}}\right).
\end{align}

The time integrations therefore reduce to the boundary contribution
$W(\infty,-\infty,\vec r)-1$, yielding the static momentum-space kernel
\begin{align}
    T(\vec q)
    = \frac{1}{N_c} \int d^3r~
    e^{i\vec q\cdot\vec r}\, 
    \tr_c \langle\infty|(w-\theta)|-\infty\rangle
    = -\frac{8\pi}{N_c} \int_0^\infty dr~
    r^2 j_0(|\vec q|r)\, \cos^2\!\left( \frac{\pi r}{2\sqrt{r^2+\rho^2}} \right).
\end{align}

Restoring the overall color normalization and the
retarded/advanced sign convention gives the static Wilson-line kernel
used in the main text.

\bibliographystyle{elsarticle-num}
\bibliography{Refs}
\end{document}